\pdfoutput=1
\PassOptionsToPackage{super,sort&compress}{natbib}
\documentclass[sn-nature,iicol]{sn-jnl}
\usepackage{xcolor}
\usepackage{graphicx}
\usepackage{placeins}

\usepackage{amsmath,amsfonts,bm}

\def\eqref#1{equation~\ref{#1}}

\def\1{\bm{1}}

\DeclareMathAlphabet{\mathsfit}{\encodingdefault}{\sfdefault}{m}{sl}
\SetMathAlphabet{\mathsfit}{bold}{\encodingdefault}{\sfdefault}{bx}{n}

\renewcommand{\eqref}[1]{\textup{(\ref{#1})}}
\usepackage{url}
\usepackage{subfigure}
\usepackage{booktabs}
\usepackage{array}
\usepackage{amsmath}
\usepackage{amssymb}
\usepackage{xspace}
\usepackage{algorithm}
\usepackage{algorithmic}
\usepackage{multirow}
\usepackage{amsthm}
\usepackage{manyfoot}
\usepackage{rotating}
\usepackage{tikz}
\usetikzlibrary{arrows.meta}
\usepackage[capitalize,nameinlink]{cleveref}
\Crefname{figure}{Fig.}{Figs.}
\Crefname{tabular}{Tab.}{Tabs.}
\Crefname{section}{\S}{\S}
\Crefname{theorem}{Thm.}{Thms.}
\Crefname{lemma}{Lemma}{Lemmas}
\Crefname{proposition}{Proposition}{Propositions}
\Crefname{corollary}{Corollary}{Corollaries}
\Crefname{algorithm}{Alg.}{Algs.}
\Crefname{example}{Ex.}{Exs.}
\Crefname{definition}{Def.}{Defs.}
\hypersetup{hypertexnames=false,bookmarksdepth=2}

\newtheoremstyle{statement}{6pt plus 4pt minus 1pt}{6pt plus 4pt minus 1pt}{\itshape}{0pt}{\bfseries}{.}{.5em}{}
\theoremstyle{statement}
\newtheorem{lemma}{Lemma}
\newtheorem{proposition}[lemma]{Proposition}
\newtheorem{corollary}[lemma]{Corollary}

\newcommand{\panel}[1]{\textbf{\sffamily #1}}

\makeatletter
\def\sectionfont{\reset@font\fontfamily{\rmdefault}\fontsize{12bp}{14bp}\bfseries\selectfont\raggedright\boldmath}
\def\subsectionfont{\reset@font\fontfamily{\rmdefault}\fontsize{10.5bp}{12.5bp}\bfseries\selectfont\raggedright\boldmath}
\renewcommand\section{\@startsection{section}{1}{\z@}{-14pt \@plus -10pt \@minus -3pt}{7pt}{\sectionfont}}
\renewcommand\subsection{\@startsection{subsection}{2}{\z@}{-11pt \@plus -8pt \@minus -2pt}{5pt}{\subsectionfont}}
\renewcommand\paragraph{\@startsection{paragraph}{4}{\z@}{-9pt \@plus -6pt \@minus -2pt}{3pt}{\paragraphfont}}
\makeatother
\title[RubriQ]{RubriQ: Rubric-Guided Group Relative Policy Optimization for Constraint-Aware Quantum Circuit Synthesis}

\author*[1,2]{\fnm{Ziqing} \sur{Guo}}\email{ziqguo@ttu.edu, zguo4@lbl.gov}

\author[1]{\fnm{Ziwen} \sur{Pan}}\email{ziwen.pan@ttu.edu}

\affil*[1]{\orgdiv{Department of Computer Science}, \orgname{Texas Tech University}, \orgaddress{\city{Lubbock}, \state{TX}, \country{USA}}}

\affil[2]{\orgname{National Energy Research Scientific Computing Center (NERSC), Lawrence Berkeley National Laboratory}, \orgaddress{\city{Berkeley}, \state{CA}, \country{USA}}}

\abstract{Quantum circuits are crucial for understanding quantum algorithms and must be correct and efficient on quantum processing units (QPUs).
Reinforcement learning (RL) with verifiers can train large language models (LLMs) to synthesize circuits; however, the reward system is unclear.
In this study, we present RubriQ, a quantum rubric, where a verifier ensures correctness and scores assess cost.
We demonstrate that, leveraging group relative policy optimization, the verifier controls correctness, while scores rank correct circuits.
Our evaluation includes unitary, statevector, stabilizer-tableau, and randomized verifiers on quantum Fourier transform (QFT), phase-estimation (QPE), and Trotter unitaries, along with Clifford, entanglement, unitary, arithmetic, and adder quantum circuits, up to 192 qubits.
Our results show that training increases correct outputs from 7\% to 91\%.
Notably, among verified circuits, the rubric closes on average 91\% of the cost gap in four of five tasks, whereas the binary reward closes less than 10\%.
We believe our framework provides valuable insights into practical quantum computing.}

\begin{document}
\maketitle

Quantum computing promises speed-ups for problems such as quantum simulation~\cite{lloyd1996universal} and linear algebra~\cite{harrow2008quantumaf}.
Algorithms are transformed into circuits when running on a quantum processor, where they are interpreted as gates, and a program can verify the circuit.
A circuit is useful if it is both accurate and cost-effective, with the cost being influenced by the machine.
In a fault-tolerant quantum computer (FTQC)~\cite{shor1996fault,tamiya2026fault} utilizing a surface code, each $T$ gate requires a magic state from distillation~\cite{bravyi2012magicstatedw,campbell2017roads} or, more recently, cultivation~\cite{rosenfeld2026magic}.
The $T$-count, or the number of $T$ gates, is a typical measure of cost~\cite{ross2016optimal,amy2019toddt,kissinger2019pyzx,kissinger2020reducing,ruiz2024alphatensorq,haenel2026tsim}.
Nevertheless, since magic states can be prepared simultaneously, both depth and qubit count also contribute to the cost.
Current noisy intermediate-scale quantum (NISQ) processing units (QPUs) impose constraints related to connectivity and native gates~\cite{preskill2018quantum,chong2017programming}.
Large language models (LLMs) are now capable of generating circuits from text~\cite{jern2025agentq,yang2024qcircuitbench}, and a program can evaluate correctness along with each of these costs.
Consequently, circuit synthesis offers a direct test of reinforcement learning with verifiable rewards (RLVR)~\cite{guo2025deepseek,shao2024deepseekmath}.

\begin{figure*}[t]
    \centering
    \includegraphics[width=\textwidth]{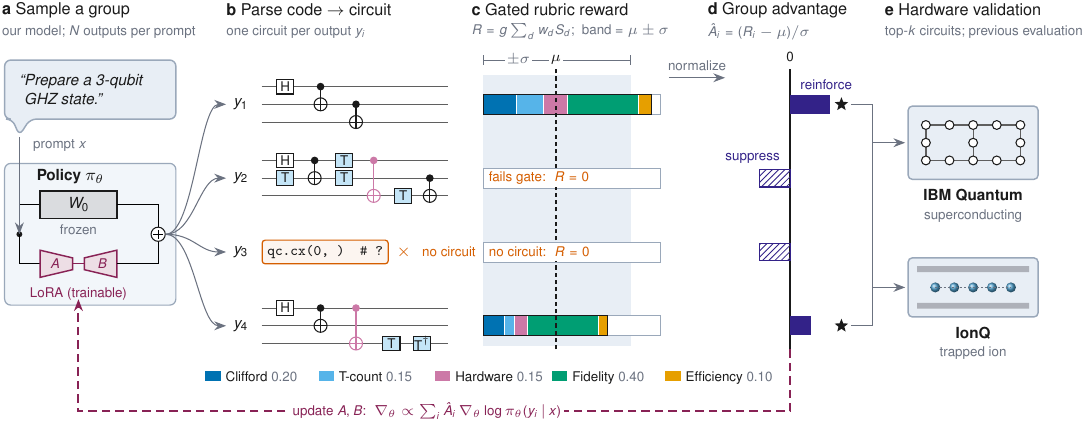}
    \caption{One RubriQ training step is depicted with illustrative values. \panel{a}, A prompt $\mathbf{x}$ is processed by the policy $\pi_\theta$, which consists of our model, with frozen weights $W_0$, equipped with trainable low-rank adaptation (LoRA) adapters $A$ and $B$~\cite{hu2022lora}. The policy generates a set of $N$ candidate solutions, with four examples shown. \panel{b}, A parser converts each candidate into a circuit; candidates that fail to parse, such as $\mathbf{y}_3$, do not produce a circuit. $T$ gates and long-range controlled-NOT (CNOT) gates are assigned a score-based color that penalizes them. \panel{c}, The correctness gate $g$ assigns a reward of zero to circuits that do not pass the verifier, exemplified by $\mathbf{y}_2$, which represents an incorrect state. Parse failures also receive $R = 0$. Circuits that pass are awarded a weighted sum of scores based on Clifford, $T$-count, hardware, fidelity, and efficiency metrics (as indicated in the legend). The two correct circuits, $\mathbf{y}_1$ and $\mathbf{y}_4$, exhibit equal fidelity but differ in cost. The band indicates the group mean $\mu_R \pm \sigma_R$. \panel{d}, Group-relative normalization provides the advantage $\hat{A}_i$ without employing a learned critic. Circuits exceeding the mean are reinforced (solid lines), while those below are suppressed (hatched lines). The gradient updates the parameters $A$ and $B$ (dashed path). \panel{e}, In a previous model evaluation, the top-$k$ circuits (indicated by stars) were executed on IBM superconducting and IonQ trapped-ion hardware.}
    \label{fig:architecture}
\end{figure*}

The task of RLVR needs a reward system that is informative and hard to exploit.
Group relative policy optimization (GRPO)~\cite{shao2024deepseekmath} learns from differences among the candidates of a prompt.
If all candidates fail an exact verifier, the prompt provides no learning signal; specifically, a single mistake in an angle or a qubit index is enough to fail.
Recently, decoupled clip and dynamic sampling policy optimization (DAPO) removes such prompts and retries~\cite{yu2025dapo}, with partial credit, a reward for the fidelity of a circuit that fails the verifier, keeping rewards informative, but it can reward shorter circuits that perform incorrect unitaries (reward hacking~\cite{skalse2022defining}).
Here, a correctness gate, which gives zero reward below a fidelity threshold, blocks this exploit but brings the first problem back.
Also, RLVR might improve only solutions that the base model already samples~\cite{yue2025does}.
A rubric grades the output with several bounded scores and considers cost only for circuits that pass the correctness gate.
Reinforcement learning (RL) trains LLMs for quantum code~\cite{yu2025quasar,dupuis2025quantum,kheiri2025qspark}, often with such staged rewards.
For example, QUASAR rewards a program in stages, from its syntax to the expectation value of its circuit~\cite{yu2025quasar}; a Qiskit code assistant uses unit tests~\cite{dupuis2025quantum}.
Rubrics also grade open-ended answers without a verifier~\cite{gunjal2025rubrics,ye2025selfrewardingrr}.
To our knowledge, no study reveals the gate and score rubric training.
We denote that RubriQ accepts any program verifier as its gate and any bounded score as a cost (\cref{fig:architecture}).

Alternatively, compilers such as Qiskit~\cite{javadi2024quantum}, TKET~\cite{sivarajah2020tketar} and CUDA-Q~\cite{cudaq} apply fixed rewrite rules to a given circuit~\cite{chong2017programming}.
RL circuit optimizers~\cite{kremer2024practicalae,fosel2021quantum} and AlphaTensor-Quantum~\cite{ruiz2024alphatensorq} start from a given circuit, and diffusion models write circuits without a reward~\cite{furrutter2024quantum}.
Agentic artificial intelligence (AI) searches for cheaper circuits under a program verifier.
Randomized tests verify kickmix circuits~\cite{khattar2026leveraging}, and a public competition scores verified circuits by their qubit and gate counts~\cite{google2026vqcc}.
Recently, LLM agents synthesized quantum programs against formal specifications in the Lean theorem prover and proved their correctness and cost~\cite{shirobokov2026discovering}.
In each of these systems, a circuit earns a cost score only if it passes the verifier.

This study demonstrates that a gated rubric functions as a two-stage reward system: the gate regulates correctness, while the cost scores determine the selection among correct circuits.
The paper presents three key contributions:
\begin{itemize}
\item \textbf{Theory.} We present four statements with proofs that are machine-checked in Lean~4 (\hyperref[sec:proofs]{Methods}). In the absence of a correct candidate, the rubric gives the same update as the binary reward, which is 1 for a circuit that passes the verifier and 0 otherwise (\cref{lem:gate}). The proportion of such groups is at least $(1 - \bar{p})^N$ for a mean pass rate $\bar{p}$ and $N$ candidates per prompt (\cref{lem:silent}). Among correct candidates, only the rubric ranks circuits by cost (\cref{prop:agree,prop:select}).
\item \textbf{Language model.} We trained the model with six rewards on our target set (\cref{tab:targets}, \hyperref[sec:results]{Results}, \cref{fig:convergence}); we refer to the base model in \hyperref[sec:base_model]{Methods}. The three gated rewards, namely the rubric, the binary reward and a scalar $T$-count reward, raise the pass rate from 0.07 (untrained) to 0.91 after 1,000 steps. Their pass rates agree within the spread between seeds throughout training (\cref{tab:mechanism}). Partial credit, a fidelity reward paid to circuits that fail the gate, stops at a pass rate of 0.76--0.77. The policy settles on near-misses, circuits with a median fidelity of 0.958 against the threshold of 0.99. We call this local optimum the near-miss trap (\hyperref[sec:r4]{Results}).
\item \textbf{Experiments.} We demonstrate that on our defined circuit families, with verifiers that scale beyond 100 qubits, the cost scores give a large improvement in the cost of correct circuits over the verifier alone in four families (\hyperref[sec:selection]{Results}, \cref{fig:selection}). Against a single-cost reward, the rubric lowers costs that the single score does not see and trades costs where it does (\cref{tab:mechanism}).
\end{itemize}

\section*{Framework design and training pipeline}
\label{sec:framework}

\subsection*{Protocol}
\label{sec:r1}
A prompt provides the target algorithm, the number of qubits and every parameter that fixes the target unitary (\cref{fig:architecture}a).
The policy, our model with trainable low-rank adaptation (LoRA) adapters~\cite{hu2022lora}, samples $N = 8$ candidates for the prompt.
A parser turns each candidate into a circuit, and a candidate that does not parse receives zero reward (\cref{fig:architecture}b).
The rubric computes five bounded scores for each circuit: the Clifford fraction $S_C$, the $T$-count score $S_T$, the hardware compatibility $S_H$, the fidelity $S_F$ and the depth efficiency $S_E$.
The reward $R$ is their weighted sum, multiplied by the correctness gate $g(c) = \mathbf{1}[F_{\mathrm{avg}} \ge 0.99]$, where $F_{\mathrm{avg}}$ is the average gate fidelity to the target unitary (\cref{fig:architecture}c, \hyperref[sec:rubric]{Methods}).

\subsection*{Training pipeline}
\label{sec:pipeline}
GRPO~\cite{shao2024deepseekmath} subtracts the group mean from each reward, divides by the group standard deviation and uses the result as the advantage in the policy gradient.
A group whose rewards are all equal has zero advantages and gives no gradient; we name it a silent group.
The gradient changes the adapter matrices $A$ and $B$, and the pretrained weights $W_0$ stay fixed (\cref{fig:architecture}a,d).
In the earlier training, the top-$k$ circuits by reward ran on IBM and IonQ processors (\cref{fig:architecture}e, \hyperref[sec:readout]{Methods}).

\subsection*{Benchmarks and verifiers}
\label{sec:benchmarks}
This study employs two benchmarks. The first benchmark is our target set (\cref{tab:targets}): quantum Fourier transform (QFT), quantum phase estimation (QPE), and Trotter targets on 2 to 8 qubits, with a held-out split based on the target unitary. Here, the precise fidelity verification requires $O(4^n)$ operations. The second benchmark includes 84 targets across five circuit families (we refer \hyperref[sec:selection_methods]{Methods}).
\begin{itemize}
\item Random Clifford operators on 5--50 qubits, the target class of RL synthesis~\cite{kremer2024practicalae}, with a stabilizer-tableau verifier~\cite{aaronson2004improved}.
\item Entanglement generation and unitary compilation, the two tasks of the diffusion-model benchmark~\cite{furrutter2024quantum}.
\item The arithmetic suite that is the standard for $T$-count optimization~\cite{amy2014polynomial,nam2018automated,ruiz2024alphatensorq}.
\item Ripple-carry adders on 12--192 qubits, verified on random basis inputs as kickmix circuits are~\cite{khattar2026leveraging}.
\end{itemize}
We emphasize that the Clifford and adder verifiers run in polynomial time, so these targets lie beyond exact simulation.

The untrained model writes code that often parses but infrequently implements the target unitary, and every target that it solves is a QPE instance (\cref{tab:mechanism}).
This pass rate sets what each reward can learn, because the gate decides which groups carry a gradient.

\begin{table}[t]
\caption{The target set of the model evaluations. Each target is one unitary, defined by one reference circuit. QFT, quantum Fourier transform; QPE, quantum phase estimation. The held-out targets are the QFT on 5 qubits, five QPE instances and the Ising chains on 4 and 5 qubits.}
\label{tab:targets}
\centering
\footnotesize
\setlength{\tabcolsep}{3pt}
\begin{tabular}{llccc}
\toprule
\textbf{Family} & \textbf{Varies over} & \textbf{Qubits} & \textbf{Targets} & \textbf{Held out} \\
\midrule
QFT & qubits & 2--8 & 7 & 1 \\
QPE & qubits, five phases $\phi$ & 3--7 & 25 & 5 \\
Trotter step & qubits, two models & 2--6 & 10 & 2 \\
\midrule
Total & & 2--8 & 42 & 8 \\
\bottomrule
\end{tabular}
\end{table}

\begin{table*}[tp]
\caption{Training statistics and the corresponding predictive statements are presented. A silent group receives eight identical rewards. A mixed group includes at least one correct and one incorrect candidate. The gated rewards consist of the rubric, the binary reward, and the gated scalar $T$-count reward. The dense variant involves the rubric with cost scores being gated. Curve values represent means over 10 iterations involving 64 candidates. The initial model evaluation employed an Open Quantum Assembly Language (OpenQASM) prompt with three seeds, while the subsequent model evaluation utilized the quantum-circuit prompt with two seeds (\hyperref[sec:training_loop]{Methods}). Mixed-group data are derived from the reward logs of rubric seeds 1 and 2, respectively. Selection data are averaged over 20 seeds with 95\% confidence intervals at step 6,000. A cost ratio is calculated relative to the minimum value among the correct candidates within a pool (\hyperref[sec:selection_methods]{Methods}). $\bar{p}$ denotes the mean pass rate as per \cref{lem:silent}; $\eta$ represents the correlation as described in \cref{prop:agree}.}
\label{tab:mechanism}
\centering
\footnotesize
\setlength{\tabcolsep}{3pt}
\begin{tabular}{>{\raggedright\arraybackslash}p{0.31\linewidth}>{\raggedright\arraybackslash}p{0.15\linewidth}>{\raggedright\arraybackslash}p{0.22\linewidth}>{\raggedright\arraybackslash}p{0.24\linewidth}}
\toprule
\textbf{Statistic} & \textbf{Reward} & \textbf{Measured} & \textbf{Predicted by} \\
\midrule
\multicolumn{4}{l}{\textit{Untrained model with the prompt of the second model evaluation, 8 samples on each target of the target set}} \\
Parse rate & --- & 0.65 & --- \\
Pass rate $\bar{p}$ & --- & 0.074 & --- \\
Targets solved at least once & --- & 16 of 42, all QPE & --- \\
\midrule
\multicolumn{4}{l}{\textit{First model evaluation, $\bar{p} = 0$}} \\
Untrained candidates that pass & --- & 0 of 336 & --- \\
Silent groups & rubric, binary & 99.9\% & all until a pass (\cref{lem:gate,lem:silent}) \\
Pass rate over the run (110--129 steps) & rubric, binary & 0.0\% & 0 until a chance pass \\
Seeds with identical candidates under both rewards & rubric, binary & 2 of 3; seed 3 until step 70 & identical until a pass (\cref{lem:gate}) \\
Informative groups & fidelity, no gate; rubric without gate & 97\% & no gate, so \cref{lem:gate} does not apply \\
Parse rate, first to last 20 steps & same & 0.45 to about 0.8 & parsing is paid \\
Fidelity of parsed circuits & same & 0.11--0.12 & --- \\
\midrule
\multicolumn{4}{l}{\textit{Second model evaluation, silent groups (\cref{fig:convergence}b)}} \\
At the start & gated & 55--65\% & $\ge (1-\bar{p})^8 = 54\%$ (\cref{lem:silent}) \\
At step 100 & gated & 40\% & --- \\
After step 400 & gated & above 90\% & all-correct groups silent (\cref{cor:idle}) \\
At the start; reaches 90\% & dense & 0\%; after step 600 & outside \cref{lem:gate} \\
\midrule
\multicolumn{4}{l}{\textit{Second model evaluation, mixed groups (seed 1; seed 2)}} \\
Mixed groups & rubric & 1,114; 691 & --- \\
Groups whose correct candidates share one reward & rubric & 1,092; 667 & $\eta = 1$ (\cref{cor:idle}) \\
Variance share $\eta^2$ of the correct/incorrect split & rubric & 0.99990; 0.99981 & 1 for one correct circuit (\cref{prop:agree}) \\
Mean root-mean-square (RMS) difference of rubric and binary advantages & rubric & 0.0010; 0.0019 & $\sqrt{2(1-\eta)}$ per group (\cref{prop:agree}) \\
Reward of a correct circuit & rubric & 0.66--0.85 & --- \\
\midrule
\multicolumn{4}{l}{\textit{Second model evaluation, pass rate (\cref{fig:convergence}a)}} \\
At steps 100; 200; 300 & gated & about 0.4; 0.72--0.75; 0.73--0.81 & equal for all gated rewards \\
At step 1,000 & gated, runs that reached it & 0.91 & --- \\
Share of Ising prompts in training & --- & 8.8\% & cap $1 - 0.088 = 0.91$ \\
Prompts solved in the last fifth: QFT and QPE; Heisenberg; Ising & gated & all; 97--99\%; none & Ising silent (\cref{lem:gate}) \\
First correct Heisenberg circuit & gated & steps 13--189 & geometric wait \\
\midrule
\multicolumn{4}{l}{\textit{Second model evaluation, near-miss trap}} \\
Parse rate at step 500 & dense; gated & 1.0; 0.83--0.93 & every parsed circuit is paid \\
Pass rate from step 300 & dense & 0.76--0.77 & share of QFT and QPE prompts, 0.765 \\
Correct Heisenberg circuits & dense & 7 of 128,000 candidates & near-miss trap \\
Heisenberg fidelity in the last fifth, median (maximum) & dense; gated & 0.958 (0.988); 1.000 & below the gate; at the gate \\
\midrule
\multicolumn{4}{l}{\textit{Cost diversity in the second model evaluation, all eight runs}} \\
Groups with two or more correct candidates & all & 38,453 & --- \\
Of these, correct candidates differ in reward & all & 238 (0.6\%) & one circuit per group (\cref{cor:idle}) \\
\midrule
\multicolumn{4}{l}{\textit{Selection experiment (\cref{fig:selection})}} \\
Pass rate & all four & 0.994--1.000 & equal below the gate (\cref{lem:gate}) \\
Steps to pass rate 0.9, $T$-count suite & gated; dense & 805--845; 890 & equal for gated rewards \\
Gap closed: Clifford; unitary; $T$-count suite; adders & rubric & $0.91 \pm 0.03$; $0.87 \pm 0.03$; $0.85 \pm 0.03$; $1.00$ & cost selection (\cref{prop:select}) \\
Same & binary & $0.05 \pm 0.11$; $0.08 \pm 0.05$; $0.04 \pm 0.09$; $-0.02 \pm 0.24$ & silent when all are correct \\
Same & scalar $T$-count & $0.05 \pm 0.11$; $0.32 \pm 0.05$; $0.68 \pm 0.04$; $1.00$ & acts through $S_T$ alone \\
Gap closed, entanglement & rubric; binary & $0.11 \pm 0.08$; $0.01 \pm 0.07$ & small spread (\cref{cor:weights}) \\
$T$-count ratio, $T$-count suite & rubric; scalar; binary & $1.02 \pm 0.01$; $1.00 \pm 0.00$; $1.26 \pm 0.04$ & uniform policy 1.31 \\
$T$-count ratio, adders & rubric; binary & $1.00$; $1.36 \pm 0.10$ & uniform policy 1.40 \\
$T$-count ratio, unitary & rubric; scalar; binary & $1.03 \pm 0.01$; $1.00 \pm 0.00$; $1.08 \pm 0.02$ & --- \\
Depth ratio: Clifford; unitary; $T$-count suite & rubric & $1.08 \pm 0.02$; $1.24 \pm 0.03$; $1.38 \pm 0.03$ & $S_E$ scores depth \\
Same & scalar $T$-count & $1.52 \pm 0.06$; $1.79 \pm 0.06$; $1.22 \pm 0.02$ & equals binary on Clifford targets \\
Same & binary & $1.52 \pm 0.06$; $1.77 \pm 0.08$; $1.25 \pm 0.02$ & --- \\
Two-qubit gate ratio: Clifford; $T$-count suite & rubric & $2.20 \pm 0.13$; $1.93 \pm 0.06$ & no score counts these gates \\
Same & scalar $T$-count & $2.06 \pm 0.14$; $1.54 \pm 0.06$ & --- \\
Same & binary & $2.06 \pm 0.14$; $1.43 \pm 0.05$ & --- \\
\bottomrule
\end{tabular}
\end{table*}

\section*{Results}
\label{sec:results_all}
\subsection*{Gate equivalence}
\label{sec:theory}

\begin{figure}[t]
\centering
\includegraphics{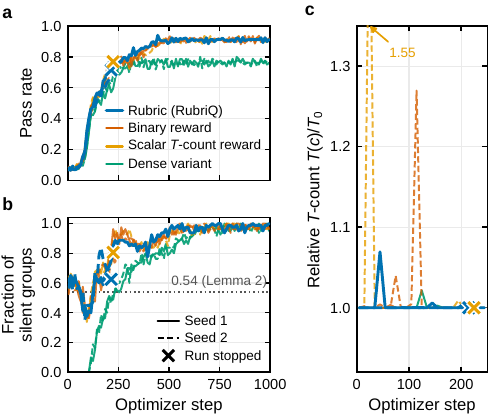}
\caption{The reward structure was evaluated under matched conditions. Four rewards were trained using the same optimizer, model, prompts, and budget, with two seeds each (represented by solid and dashed lines). The curves represent means over 10 steps of 64 candidates. In panel (a), the fraction of candidates passing the correctness gate on the training prompts is depicted. The three gated rewards align (\cref{lem:gate,cor:idle}), while the dense variant halts at the proportion of QFT and QPE prompts. Panel (b) illustrates the fraction of silent groups, where eight rewards are equal and provide no gradient. Initially, the gated rewards are slightly above the bound $(1-\bar{p})^8 = 0.54$ as indicated in \cref{lem:silent} (dotted line). Panel (c) shows the $T$-count of admissible candidates relative to the reference circuit of their target, from steps 0 to 250. The ratio remains at 1.00 in every subsequent bin (\cref{tab:main}). Notably, one scalar seed-2 bin achieves a ratio of 1.55 (indicated by an arrow). Six runs extended to 1,000 steps; however, the rubric seed-2 and scalar seed-1 runs concluded at steps 216 and 229, respectively (marked by crosses). All values underlying the curves are detailed in \cref{tab:mechanism}.}
\label{fig:convergence}
\end{figure}

The correctness gate multiplies the reward by $g(c) \in \{0, 1\}$. Consequently, an incorrect circuit receives a reward of $R = 0$, regardless of its other four scores, as does a candidate that fails to parse. The rubric and the binary reward system are therefore consistent for every candidate that does not succeed. We note that the proof of each statement below is machine-checked in Lean~4 (\hyperref[sec:proofs]{Methods}).

\begin{lemma}\label{lem:gate}
Let $R = g\,f$, where $g$ is the correctness gate and $f$ is any bounded function of the circuit.
In a group in which no candidate passes the gate, every advantage $\hat{A}_i$ is zero under $R$ and under the binary reward $g$.
\end{lemma}

Because \Cref{lem:gate} is derived from the factor $g$, this ensures the validity for any number and any weights of the scores. The loss function lacks a Kullback--Leibler (KL) term (\hyperref[sec:training_loop]{Methods}), resulting in a zero gradient for a step where all groups are silent. Here, the $N$ candidates within a group are independent samples from $\pi_\theta(\cdot \mid \mathbf{x})$. Each candidate passes through the gate with a probability $p_{\mathbf{x}}$ associated with its prompt, and these probabilities determine the proportion of silent groups.

\begin{lemma}\label{lem:silent}
Let $\bar{p} = \mathbb{E}_{\mathbf{x}}[p_{\mathbf{x}}]$ be the mean pass rate over prompts.
Under any gated reward, the expected share of silent groups is at least $\mathbb{E}_{\mathbf{x}}[(1 - p_{\mathbf{x}})^N] \ge (1 - \bar{p})^N$.
\end{lemma}

At $\bar{p} = 0$, all groups remain silent, and no gated reward can be initiated. This scenario was observed in the first model evaluation. Here, under an OpenQASM prompt, the untrained model did not select any candidate. Under the rubric and binary reward system, nearly all groups remained silent, resulting in a pass rate of zero (\cref{tab:mechanism}). Two dense rewards without gating maintained informative groups and conveyed the output format, although they did not impart the target unitary.

In the second model evaluation, $\bar{p}$ was increased above zero using the quantum-circuit prompt, which lets the policy build the circuit with the Qiskit package (\hyperref[sec:state_action]{Methods}). Consequently, the silent proportion of the three gated rewards began just above the threshold specified in \cref{lem:silent} (\cref{fig:convergence}b, \cref{tab:mechanism}). This proportion decreased as training elevated $p_{\mathbf{x}}$, but subsequently increased from the opposite direction. A well-crafted prompt results in groups where every candidate is correct, rendering the group silent under the binary reward. It also remains silent under the rubric when the correct candidates form a single circuit (\cref{cor:idle}). The dense variant assesses fidelity below the gate and is not constrained by \cref{lem:gate}, thus it initially had no silent group.

In a heterogeneous group comprising at least one correct and one incorrect candidate, the gate assigns a reward of zero to each incorrect candidate. The rubric diverges from the binary reward by distributing $R$ among the correct candidates.

\begin{proposition}\label[proposition]{prop:agree}
Let $\hat{A}^{R}$ and $\hat{A}^{B}$ be the advantage vectors of a mixed group under the rubric and under the binary reward, with $\varepsilon = 0$.
Then $\tfrac{1}{N}\lVert \hat{A}^{R} - \hat{A}^{B} \rVert^2 = 2(1 - \eta)$, where $\eta$ is the correlation between the two reward vectors.
Here $\eta^2$ is the share of the variance of $R$ that the split into correct and incorrect candidates explains.
It equals 1 exactly when all correct candidates have the same reward.
\end{proposition}

\begin{corollary}
\label[corollary]{cor:idle}
If the correct candidates of a group share one reward $f^\ast$, then $\hat{A}^{R} = \hat{A}^{B}$ in that group.
This holds when they are one circuit.
If all candidates of the group are correct, the group is silent under both rewards.
\end{corollary}

In the training logs of the rubric, the correct candidates within a mixed group consistently received one reward in nearly every instance (\cref{tab:mechanism}). The equal distribution of rewards across all five scores indicates the presence of a single circuit per group.

\subsection*{Training dynamics}
\phantomsection\label{sec:results}
\Cref{lem:gate} and \cref{cor:idle} confer identical advantages to both the rubric and the binary reward across all groups, with the exception of mixed groups containing distinct correct circuits. Such groups are infrequent, and \cref{prop:agree} quantifies their impact as $2(1 - \eta)$, where \cref{cor:weights} provides the value of $\eta$ for any given weights. The gated scalar $T$-count reward is expressed as $g\,S_T$, and thus, the same conclusions are applicable. Consequently, the pass rates of the three gated rewards are consistent within the variability observed between seeds (\cref{fig:convergence}a, \cref{tab:mechanism}).

We observe that a Heisenberg prompt exhibits a low pass probability, denoted as $p_{\mathbf{x}}$, under the untrained policy. The probability that a group of such prompts contsains a correct circuit is given by $1 - (1 - p_{\mathbf{x}})^N$. Consequently, the number of groups required until the first correct circuit is identified follows a geometric distribution. This circuit subsequently demonstrates a positive advantage, with each update increasing $p_{\mathbf{x}}$ and reducing the subsequent waiting time. Due to the wide variance inherent in a geometric distribution, the first correct Heisenberg circuit emerged at significantly different steps across various runs (\cref{tab:mechanism}).

\subsection*{Partial credit}
\phantomsection\label{sec:r4}
A quantum circuit is considered below the gate when its fidelity satisfies the condition $F_{\mathrm{avg}} < \tau_F = 0.99$; otherwise, we refer at the gate. In the dense variant, partial credit is represented by the fidelity term $w_F S_F$, which is awarded to every parsed circuit, while cost scores are incurred only at the gate. Here, two incorrect circuits with differing fidelities receive distinct rewards, rendering \cref{lem:gate} inapplicable and allowing a group without a correct circuit to still produce a gradient. This mechanism enables the policy to learn prior to sampling its first correct circuit. Specifically, a parsed circuit accrues more than an unparsed one, prompting the policy to consistently return a parsable circuit for each prompt (\cref{tab:mechanism}).

We define a near-miss as a parsed circuit whose fidelity approaches but does not meet the threshold $\tau_F$. Under the partial credit system, a near-miss is awarded close to the full fidelity term $w_F S_F$, despite implementing an incorrect unitary. In a group devoid of correct circuits, the near-miss consequently receives the highest reward and a positive advantage, leading to an increased probability in its favor. In policy-gradient training, a positive covariance between the log-probability of a candidate and its advantage reduces the policy's entropy~\cite{cui2025entropy}. Hence, a near-miss that exceeds the group mean results in a positive covariance. This situation allows the policy to focus on near-misses, reducing the diversity of sampled circuits and decreasing the likelihood of sampling a correct circuit.

We note that the near-miss trap as the resulting local optimum, in which the policy places most of its probability on near-misses of a prompt and rarely samples a correct circuit for that prompt. Under a gated reward, the same groups are silent, so that the policy keeps its sampling distribution until a correct circuit appears. Partial credit thus retains the global optimum and adds a local optimum on the path of gradient ascent. In the second model evaluation, the Heisenberg circuits of the dense variant remained below the gate, with a median fidelity of 0.958 and a maximum of 0.988, whereas those of the gated rewards reached it (\cref{tab:mechanism}). Consequently, the pass rate of the dense variant is limited to the proportion of QFT and QPE prompts (\cref{fig:convergence}a).

\subsection*{Cost selection}
\phantomsection\label{sec:selection}

\begin{figure}[t]
\centering
\includegraphics{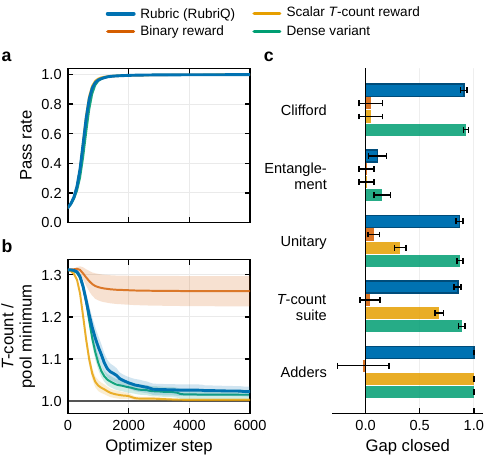}
\caption{Cost selection in the selection experiment involves a softmax policy applied to a pool of verified circuits, which receives the GRPO update under four distinct rewards. The lines represent means over 20 seeds, while the bands and error bars denote 95\% confidence intervals. In \panel{a}, the pass rate on the 17 targets of the $T$-count suite is depicted, showing that all rewards achieve correctness at a consistent rate. \panel{b} illustrates the $T$-count of the correct candidates under the policy, relative to the lowest $T$-count among the correct candidates in each pool. Notably, the binary reward remains close to the uniform policy at 1.31. \panel{c} demonstrates the gap closure at step 6,000 across the five families (\hyperref[sec:selection_methods]{Methods}). It is observed that every $T$-count is zero for the Clifford and entanglement targets, resulting in the scalar reward equating to the binary reward in these cases. The benchmark comprises 84 targets and 9,850 candidates, with values detailed in \cref{tab:mechanism}.}
\label{fig:selection}
\end{figure}

Continuing on \Cref{lem:gate}, the cost scores based on a group with two or more correct candidates.

\begin{proposition}\label[proposition]{prop:select}
Let $R = g\,f$, and let candidates $i$ and $j$ of a group pass the gate.
Then $\hat{A}^{R}_i - \hat{A}^{R}_j = (f_i - f_j)/(\sigma_R + \varepsilon)$ and $\hat{A}^{B}_i - \hat{A}^{B}_j = 0$.
If every candidate of the group passes, the group is silent under the binary reward.
Under $R$ its advantages are then the standardized scores $(f_i - \mu_f)/(\sigma_f + \varepsilon)$.
\end{proposition}

A gated rubric operates in two distinct stages. In instances where correct candidates are infrequent, the gate establishes the advantages, and the rubric subsequently applies a binary reward (\cref{prop:agree}). When a group contains multiple correct candidates, the binary reward is insufficient for ranking, necessitating the rubric to rank them based on cost. The second stage requires distinct correct circuits within a single group.

The selection experiment provides distinct correct circuits and maintains the update. For each target, a pool contains correct circuits derived from published constructions and compilers, alongside nine times as many incorrect circuits (\hyperref[sec:selection_methods]{Methods}). A softmax policy over the pool substitutes for the model. This policy initially starts uniformly and is updated according to Eq.~\eqref{eq:grpo_gradient}.
All four rewards reached a pass rate above 0.99.
The three gated rewards crossed a pass rate of 0.9 within one logging interval of each other (\cref{fig:selection}a, \cref{tab:mechanism}).
The rewards then separated in cost.
We note that the binary reward is the control, with no cost term.
The scalar $T$-count reward is the baseline, with one cost score behind the same gate.
The gap closed measures the mean reward $R$ of correct candidates; it is in units of $R$ and thus favours the rubric, so we also compare gate counts (\cref{tab:mechanism}).

The scalar reward achieved the minimum $T$-count for each pool (\cref{fig:selection}b). The rubric reduced the depth in instances where the scalar reward did not. Specifically, for Clifford and unitary targets, the depth was 8\% and 24\% higher than the minimum, compared to 52\% and 79\% lower under the scalar reward. In the $T$-count suite, the order was reversed, where the rubric was 38\% lower, whereas the scalar reward was 22\% lower. Consequently, various scores identify instances where a single cost does not differentiate the circuits, and they exchange costs where differentiation occurs.

\subsection*{Compression and compiler baselines}
\label{sec:r5}

\begin{table*}[t]
\caption{The evaluation of correctness and $T$-count was conducted on the target set. Compression is defined as $T_0/T(c)$ for outputs that implement the target unitary, with $F_{\mathrm{avg}} \ge 0.99$. Compilers were executed once on the reference circuit for each target. Policies were sampled eight times per held-out target, resulting in $n = 64$ candidates. Correct outputs refers to the count of admissible candidates, while solved indicates the number of held-out targets with at least one admissible candidate. Every correct held-out circuit of the two binary policies (81) and of rubric seed 1 (36) utilized Qiskit's library QFT. The two unsolved held-out targets are the Ising chains on 4 and 5 qubits, which no policy managed to solve during training. Notably, the run was halted at step 216 and was evaluated at its step-200 checkpoint.}
\label{tab:main}
\centering
\footnotesize
\setlength{\tabcolsep}{4pt}
\begin{tabular}{llccc}
\toprule
\textbf{Method} & \textbf{Run} & \textbf{Correct outputs} & \textbf{Compression $T_0/T$} & \textbf{Held-out solved} \\
\midrule
\multicolumn{5}{l}{\textit{Compilers (one run per target)}} \\
PyZX, ZX-calculus full reduction & target set & 42 / 42 & 1.30 & --- \\
TKET, full peephole & target set & 42 / 42 & 1.13 & --- \\
TODD; PyZX phase-block & 4 Clifford+$T$ targets & 4 / 4 & 1.25 & --- \\
Qiskit, level 3 & target set & 42 / 42 & 0.61 & --- \\
BQSKit, level 2 & target set & 42 / 42 & 0.36 & --- \\
\midrule
\multicolumn{5}{l}{\textit{Policies (8 samples per held-out target; step-1,000 checkpoint unless stated)}} \\
Base model, no training & --- & 2 / 64 & 1.00 & 2 / 8 \\
Rubric (RubriQ) & seed 1 & 36 / 64 & 1.00 & 6 / 8 \\
Rubric (RubriQ) & seed 2, step 200$^*$ & 40 / 64 & 1.00 & 6 / 8 \\
Binary reward & seed 1 & 38 / 64 & 1.00 & 6 / 8 \\
Binary reward & seed 2 & 43 / 64 & 1.00 & 6 / 8 \\
Scalar $T$-count reward & seed 1, step 200$^*$ & 29 / 64 & 1.00 & 5 / 8 \\
Scalar $T$-count reward & seed 2 & 45 / 64 & 1.00 & 6 / 8 \\
Rubric, cost scores gated & seed 1 & 36 / 64 & 1.00 & 6 / 8 \\
Rubric, cost scores gated & seed 2 & 43 / 64 & 1.00 & 6 / 8 \\
\bottomrule
\end{tabular}
\end{table*}

The sandbox facilitates the invocation of quantum circuits, with the estimator quantifying their library decomposition (\hyperref[sec:state_action]{Methods}). Invoking the library QFT replicates the reference circuit of a QFT target and the concluding block of a QPE target. The gate accommodates such an invocation, thereby increasing its probability compared to manually written circuits~\cite{yue2025does}. This phenomenon is known as shortcut learning~\cite{geirhos2020shortcut}, where the policy achieves the reward by reusing the reference implementation, a strategy not intended by the task designer. The $T$-count is inherently $T_0$, and all correct circuits exhibit this count (\cref{fig:convergence}c, \cref{tab:main}). Consequently, the correct candidates within a group converge to a single circuit, and \cref{cor:idle} results in cost scores that do not differentiate among them.

The library call is transferred to an unobserved QFT or QPE target without modification, enabling the trained policies to address the reserved QFT and QPE targets (\cref{tab:main}). However, they do not resolve any reserved Ising target, as \cref{lem:gate} provided those prompts with no gradient during training. An approximate QFT excludes its smallest controlled rotation while maintaining high fidelity~\cite{coppersmith2002approximate,barenco1996approximate}, and it continues to function effectively on five or more qubits. 

We employ Qiskit, TKET, PyZX, TODD, and BQSKit as distinct tools on the reference circuit for each target (\hyperref[sec:baselines]{Methods}). Each compiler output successfully passes the gate (\cref{tab:main}). ZX-calculus rewriting consolidates phase rotations acting on the same parity of qubits, thereby eliminating $T$ gates~\cite{kissinger2020reducing}. Qiskit's level-3 pass and BQSKit's resynthesis substitute gate blocks with arbitrary-angle rotations. The estimator assigns a charge of 50 $T$ gates for each such rotation, following the Ross--Selinger count with a precision of approximately $10^{-5}$~\cite{ross2016optimal} (\hyperref[sec:rubric]{Methods}). Consequently, these two tools increase the $T$-count compared to the reference. ZX-calculus and peephole rewriting outperform every trained policy in the metric that the rubric was designed to reward.

\subsection*{Hardware}
\label{sec:r6}

\begin{figure}[t]
\centering
\includegraphics[width=\columnwidth]{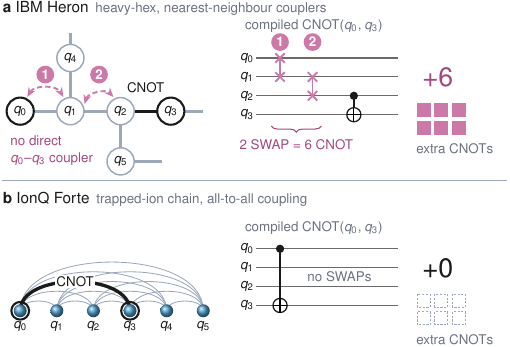}\\[4pt]
\includegraphics{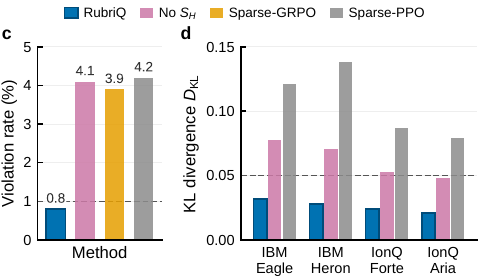}
\caption{Routing cost and executability on near-term hardware. \panel{a}, On the heavy-hex lattice of IBM Heron, two SWAP gates (1, 2), each comprising three CNOT gates, position $q_0$ adjacent to $q_3$. \panel{b}, On the trapped-ion chain of IonQ Forte, every pair of qubits is inherently coupled, thus eliminating the need for a SWAP gate for the same CNOT operation. The SWAP overhead is accounted for by $S_H$ (\hyperref[sec:rubric]{Methods}). \panel{c}, The hardware constraint violation rate $\mathcal{V}$ from the previous model evaluation is depicted, with the dashed line indicating the 1\% target. \panel{d}, The Kullback--Leibler divergence between the measured and noiseless output distributions of a 4-qubit QFT circuit, with 4,096 shots per circuit, is shown. The divergence for RubriQ (numerical values) remains below the threshold $D_{\mathrm{KL}} = 0.05$. The earlier model evaluation reported no divergence for Sparse-GRPO. In \panel{c} and \panel{d}, each bar represents a reported value, without error bars, due to the absence of archived logs from the earlier model evaluation. The policies were trained using the previous evaluator.}
\label{fig:hardware}
\end{figure}

The current processor imposes constraints on the number of qubits, the native gate set, and the coupling map. Implementing a two-qubit gate between uncoupled qubits necessitates the use of SWAP gates, with each SWAP gate introducing three CNOT gates and their associated errors (\cref{fig:hardware}a,b). The parameter $S_H$ accounts for gates outside the native set and the estimated SWAP overhead (\hyperref[sec:rubric]{Methods}), thereby prioritizing circuits that can be executed by a backend as originally designed. A previous model evaluation assessed this effect on IBM and IonQ processors, and these experiments were not replicated (\cref{fig:hardware}c,d).

Readout errors contribute to gate errors in every measured distribution. For independent readout errors per qubit~\cite{bravyi2021mitigating}, a symmetric error results in the uniform output of the Quantum Fourier Transform (QFT) test circuit mapping onto itself. The asymmetry between the two flip probabilities of each qubit increases the Kullback-Leibler divergence ($D_{\mathrm{KL}}$) beyond its pre-readout value, as bounded by Proposition \ref{prop:readout} (\hyperref[sec:readout]{Methods}). In the absence of $S_H$, the violation rate was 4.1\%, compared to 0.8\% for RubriQ. 

\section*{Discussion}
\label{sec:discussion}
This study demonstrates that a gated rubric functions as a two-stage reward system. The initial stage involves the gate, which determines the beginning of learning and the rate at which correctness improves, aligning the rubric with the binary reward (\cref{lem:gate}). Subsequently, the cost scores dictate which correct circuit the policy retains (\cref{prop:select}). The model evaluation validate the first stage, as three gated rewards produced a single learning curve. The selection experiment corroborates the second stage across five circuit families. Furthermore, the library shortcut elucidates why the second stage remained inactive during the evaluation (\cref{cor:idle}). We also denote that partial credit lowers the entropy of the policy below the gate~\cite{cui2025entropy}.

Indeed, the rubric scores are interchangeable; for instance, in the $T$-count suite, the rubric sacrificed depth, which the scalar reward preserved. A cost score requires distinct correct circuits within a single category. This can be achieved by using a prompt that prohibits library calls or by setting a target with flexibility, such as the Quantum Fourier Transform (QFT) on five or more qubits. We note that a substantial fidelity weight shifts the rubric towards a binary reward system. This adjustment increases the average reward for correct circuits without affecting their variability (\cref{cor:weights}).

Meanwhile, note that verification decides how far the benchmark scales.
An exact fidelity check costs $O(4^{n})$ operations.
A stabilizer tableau verifies a Clifford circuit in polynomial time.
A classical simulator runs a kickmix circuit on a basis input in time linear in its gate count~\cite{khattar2026leveraging}.
Randomized tests on such inputs bound the error of the circuit, and our adder targets use this protocol up to 192 qubits.
By \cref{lem:gate}, we emphasize that any verifier, such as a statevector simulator or a tensor-network contraction, serves as an equivalence checker.
As such, the recent competition~\cite{google2026vqcc} offers broader benchmarking.

\section*{Methods}
\label{sec:method}

\subsection*{Problem statement}
\label{sec:problem_statement}
A prompt $\mathbf{x}$ names a target algorithm $\mathcal{A}$ in natural language, for example a QFT on $n$ qubits.
The policy $\pi_\theta$ writes circuit code $\mathbf{y}$, and a program parses it into a circuit $c = \mathrm{parse}(\mathbf{y})$.
We seek the policy that maximizes the expected reward
\begin{equation}
\label{eq:objective}
\max_\theta \; \mathbb{E}_{\mathbf{x}\sim\mathcal{D},\, \mathbf{y}\sim\pi_\theta(\cdot\mid\mathbf{x})}
\Big[ R(\mathbf{y}; \mathcal{A}) \Big],
\end{equation}
where $\mathcal{D}$ is the prompt dataset and $R$ is the reward defined below.
Correctness is a gate, not a trade-off.
A circuit is admissible when its average gate fidelity to the target unitary satisfies $F_{\text{avg}}(U_{\mathbf{y}}, U_{\mathcal{A}}) \ge \tau_F$ with $\tau_F = 0.99$.
An inadmissible circuit receives zero reward, and we report FTQC cost for admissible circuits.
Supplementary Note~1 lists all symbols.

\subsection*{Group relative policy optimization}
\label{sec:grpo}
GRPO~\cite{shao2024deepseekmath} removes the value critic of proximal policy optimization (PPO)~\cite{schulman2017proximalpo}.
For each prompt it samples a group of $N$ candidates and normalizes each reward by the group statistics:
\begin{align}
\label{eq:grpo_advantage}
\hat{A}_i &= \frac{R_i - \mu_R}{\sigma_R + \varepsilon}, \qquad
\mu_R = \frac{1}{N}\sum_{j=1}^{N} R_j, \\
\sigma_R &= \sqrt{\frac{1}{N}\sum_{j=1}^{N}(R_j - \mu_R)^2},
\end{align}
where $R_i$ is the reward of the $i$-th candidate and $\varepsilon$ is a small constant.
The policy gradient is
\begin{equation}
\label{eq:grpo_gradient}
\nabla_\theta J(\theta) = \frac{1}{N}\sum_{i=1}^{N}
\nabla_\theta \log \pi_\theta(\mathbf{y}_i \mid \mathbf{x})\, \hat{A}_i,
\end{equation}
where $\mathbf{y}_i$ is the $i$-th candidate.
The update uses the raw rewards of one group, so it needs no critic and no bootstrapped value targets.
A silent group has $\sigma_R = 0$, and we report the fraction of silent groups per step.

\subsection*{Targets and benchmark}
\label{sec:datasets}
Each target is defined by one explicit reference circuit, and the target unitary is the unitary of that circuit in Qiskit's little-endian convention.
The target set has 42 targets in three families (\cref{tab:targets}):
\begin{itemize}
\item QFT on 2--8 qubits, defined by Qiskit's \texttt{QFTGate} with its final swaps (7 targets)~\cite{coppersmith2002approximate}.
\item QPE on 3--7 qubits for the phase gate $P(2\pi\phi)$ with $\phi \in \{1/8, 1/4, 3/8, 1/2, 3/4\}$ (25 targets)~\cite{kitaev1995quantum}.
The evaluation qubits are $q_0 \ldots q_{m-1}$ with $m = n - 1$, and the target qubit is $q_m$.
The circuit is $H^{\otimes m}$, then controlled $P(2\pi\phi\,2^k)$ from $q_k$, then the inverse \texttt{QFTGate}$(m)$.
\item One first-order Trotter step of $e^{-iHt}$ for the Heisenberg XXX chain and the transverse-field Ising chain on 2--6 qubits (10 targets)~\cite{lloyd1996universal}. Here $J = h = t = 1$, and the prompt states the bond order.
\end{itemize}
The reference $T$-count $T_0$ of a target is the rubric's own estimate on the reference circuit, expanded into basic gates.
We exclude 8 targets based on target unitary (20\%, seed 42): Quantum Fourier Transform (QFT) on 5 qubits, five instances of Quantum Phase Estimation (QPE), and the Ising chain on 4 and 5 qubits. Each excluded target belongs to the same task family as the training targets but differs in its unitary. The data-encoding tasks from the previous model evaluation are not included in this benchmark, as state preparation lacks a target unitary. The previous model evaluation's benchmark of 1,500 circuits and its logs are absent from the code archive.

\subsection*{Prompting, generation and circuit extraction}
\label{sec:state_action}
Each prompt has two parts in the chat format of the base model.
The system prompt states the task constraints, names the qubit-order convention, and shows the two accepted output formats with a four-gate example each: OpenQASM~2.0 in a \texttt{qasm} code block, or a circuit written with the Qiskit package in a \texttt{python} code block that builds a \texttt{QuantumCircuit}.
The user prompt names the target, the qubit count and every parameter that fixes the unitary, for example the phase $\phi$ and the bond order.
The full input is $\mathbf{x} = [\mathbf{x}_{\text{sys}};\, \mathbf{x}_{\text{user}}]$.
Supplementary Note~5 gives both prompts in full.

The policy generates a candidate sequence $\mathbf{y}$ on a token-by-token basis, employing a temperature of 0.8 and nucleus sampling~\cite{holtzman2019curious} with a top-$p$ value of 0.95, for a maximum of 1,024 tokens. Each complete candidate sequence constitutes a single action, and we generate $N = 8$ candidate sequences per prompt at each step.

A parser extracts the circuit with five strategies in order:
\begin{enumerate}
\item Parse the text as OpenQASM~2.0 if it begins with the \texttt{OPENQASM} header.
\item Extract a \texttt{qasm} code block and proceed with parsing.
\item Address common deviations within that block and reparse. These deviations include a missing header, OpenQASM~3 register declarations, call-style gate syntax such as \texttt{cu1(pi/4, q[1], q[0])}, gate aliases, and missing commas.
\item Extract a \texttt{python} code block and execute it within a sandbox environment. Imports are statically checked and restricted to Qiskit's circuit, library, and \texttt{quantum\_info} modules, as well as NumPy and \texttt{math}. The code is devoid of file, network, or process access, and is subject to a 5-second execution limit. Methods deprecated in Qiskit~1.0, such as \texttt{cu1} and \texttt{u3}, are mapped to their current equivalents. The largest \texttt{QuantumCircuit} remaining in the namespace is identified as the candidate. Calls to Qiskit's library circuits, such as \texttt{QFT}, are permissible; the estimator accounts for the library decomposition, which serves as the reference circuit.
\item Parse the raw text as OpenQASM~2.0.
\end{enumerate}
A candidate that yields no circuit receives $R = 0$.
We call the prompt with these two formats the quantum-circuit prompt. With it, 93\% of base-model outputs build the circuit with the Qiskit package.

\phantomsection\label{sec:base_model}The foundational model employed is Qwen2.5-Coder-7B-Instruct~\cite{hui2024qwen2}, an instruction-tuned code model comprising 7 billion parameters. We implement LoRA adapters~\cite{hu2022lora} with a rank of $r = 64$ and $\alpha = 128$ on the query and value projections of each attention layer, while maintaining the pretrained weights in a frozen state. For a projection with a pretrained weight matrix $W_0$ and input $\mathbf{h}$, the adapted layer computes $W_0 \mathbf{h} + \tfrac{\alpha}{r} B A\, \mathbf{h}$. The matrix $A$ consists of $r$ rows, and $B$ comprises $r$ columns, resulting in the product $BA$ having a rank of at most $r$. Initially, $B$ is set to zero, thereby aligning the policy with the base model. The parameters $\theta$ in Eq.~\eqref{eq:grpo_gradient} correspond to the entries of all matrices $A$ and $B$; training modifies these parameters while keeping each $W_0$ constant. The adapters encompass 20 million parameters, representing 0.26\% of the entire model.

\subsection*{Rubric}
\label{sec:rubric}
We give each parsed circuit five scores $S_d \in [0,1]$ and combine them with fixed weights:
\begin{align}
R(\mathbf{y}) &= g(c) \big[ w_F S_F(c) + R_{\mathrm{FT}}(c) + R_{\mathrm{NT}}(c) \big], \label{eq:rubric_reward}\\
R_{\mathrm{FT}}(c) &= w_C S_C(c) + w_T S_T(c) + w_E S_E(c), \nonumber\\
R_{\mathrm{NT}}(c) &= w_H S_H(c), \nonumber
\end{align}
where $g(c) = \mathbf{1}[F_{\text{avg}}(c) \ge \tau_F]$ is the correctness gate.
$R_{\mathrm{FT}}$ is the FTQC cost block for a surface-code Clifford+$T$ architecture.
$R_{\mathrm{NT}}$ is the hardware compatibility block for the IBM and IonQ backends.
The reward is zero if the candidate does not parse or if the circuit fails the gate.
The weights are $(w_C, w_T, w_H, w_F, w_E) = (0.20, 0.15, 0.15, 0.40, 0.10)$.
They are non-negative and sum to one, so $R$ is a convex combination of the five scores and $R \in [0,1]$.
Fidelity has the largest weight because a cheap circuit that implements the wrong unitary has no value.
We chose the weights by a grid search in the earlier model evaluation.

\paragraph{Clifford fraction $S_C$}
We classify each gate exactly with Qiskit's \texttt{Clifford} class~\cite{aaronson2004improved,gottesman1998heisenberg}.
The score is the Clifford fraction of the operational gates:
\begin{equation}
S_C(c) = \frac{|\{g \in c : g \text{ is Clifford}\}|}{|\{g \in c : g \text{ is operational}\}|}.
\end{equation}

\paragraph{$T$-count $S_T$}
The estimator assigns a count $T(g)$ to each gate and aggregates these counts as $T(c) = \sum_{g \in c} T(g)$. Clifford gates contribute a value of 0. The $T$ gate contributes a value of 1, while the Toffoli and Fredkin gates each contribute 7, based on their standard ancilla-free decompositions~\cite{amy2012ama,selinger2013quantum}. Any other gate is decomposed into CNOT (CX) and $R_z$ gates. Each $R_z(\theta)$ contributes 0 for a Clifford angle, 1 for an odd multiple of $\pi/4$, and 50 otherwise. The constant 50 is derived from the Ross--Selinger bound $3\lceil\log_2(1/\varepsilon)\rceil$~\cite{ross2016optimal}, with a synthesis precision of $\varepsilon \approx 10^{-5}$, resulting in $3 \times 17 = 51$. At $\varepsilon = 10^{-10}$, the same bound yields $3 \times 34 = 102$, although the estimator does not utilize this value. The estimator also defines $T_0$, incorporating the constant into both $T_0$ and $T(c)$. The score is calculated relative to the reference as follows:\begin{equation}\label{eq:st}S_T(c) = \exp\!\Big(-\ln 2 \cdot \frac{T(c)}{T_0}\Big),\end{equation} such that $S_T = 1$ when $T(c) = 0$, $0.5$ when $T(c) = T_0$, and $0.25$ when $T(c) = 2T_0$. 

\paragraph{Hardware compatibility $S_H$}
We check the circuit against each backend $b$ in $\{$IBM Eagle, IBM Heron, IonQ Forte, IonQ Aria$\}$ and keep the best score:
\begin{equation}
\begin{aligned}
S_H(c) = \max_b \Big[ & w_q \cdot \mathbf{1}[n_q(c) \le Q_b] \\
& + w_d \cdot \mathbf{1}[D(c) \le D_b] \\
& + w_n \cdot \text{NativeFrac}(c, b) \\
& + w_r \cdot \exp(-\lambda \cdot \text{SwapOH}(c, b)) \Big],
\end{aligned}
\end{equation}
where $Q_b$ and $D_b$ are the qubit and depth limits of $b$.
The variable $\text{NativeFrac}$ represents the proportion of gates within the native set of $b$, while $\text{SwapOH}$ denotes the estimated SWAP overhead. The assigned weights are $(w_q, w_d, w_n, w_r) = (0.3, 0.2, 0.3, 0.2)$. The routing term employs $\lambda = 0.3$ to evaluate the ratio of $\text{SwapOH}$ to the number of two-qubit gates. For a backend featuring all-to-all connectivity, $\text{SwapOH}$ equals 0. Conversely, for a backend with a restricted topology, each two-qubit gate between qubits separated by an index distance $\delta > 1$ incurs an additional $\delta - 1$ SWAPs. The metric $S_H$ assesses whether a circuit is executable on current hardware.

\paragraph{Fidelity $S_F$}
We compute the process fidelity from the Hilbert--Schmidt inner product and convert it to the average gate fidelity:
\begin{equation}
\label{eq:fidelity}
\begin{aligned}
F_{\text{process}} &= \frac{|\text{Tr}(U_{\mathcal{A}}^\dagger \, U_c)|^2}{d^2}, \\
F_{\text{avg}} &= \frac{d \cdot F_{\text{process}} + 1}{d + 1},
\end{aligned}
\end{equation}
where $d = 2^n$, $U_{\mathcal{A}}$ is the target unitary and $U_c$ is the circuit unitary after we strip measurements and resets.
The target width $n$ is specified in the prompt; therefore, a circuit with an incorrect number of qubits receives a score of $F_{\text{avg}} = 0$. We also consider the target in a fully reversed qubit order as acceptable. We compute $U_c$ precisely, with all targets comprising a maximum of 8 qubits.

\paragraph{Correctness gate}
The fidelity is incorporated into the reward in two ways: as the gate $g(c)$ with $\tau_F = 0.99$, and as the weighted term $w_F S_F$. A circuit that does not pass the gate receives a reward of $R = 0$, equivalent to a parse failure, regardless of its other scores. The gate effectively closes the reward-hacking channel, where the policy reduces the circuit by altering the target unitary. Additionally, it implies that a group with no successful candidates exhibits zero variance, similar to a binary reward (\cref{lem:gate}). 

\paragraph{Efficiency $S_E$}
A depth-normalized score rewards compact circuits:
\begin{equation}
S_E(c) = \exp\!\Big(-\frac{\beta \cdot D(c)}{n_q^2}\Big),
\end{equation}
where $D(c)$ is the circuit depth and $\beta = 0.5$.

\paragraph{Dependence on the weights}
By \cref{prop:agree}, the rubric differs from the binary reward through the correlation $\eta$, and $\eta$ has a closed form in the weights.
Let $\mathbf{w} = (w_C, w_T, w_H, w_F, w_E)$, and let $\mathbf{s}_i$ be the vector of the five scores of candidate $i$ in the same order.
A correct candidate then has the reward $\mathbf{w}^\top \mathbf{s}_i$.

\begin{corollary}\label[corollary]{cor:weights}
Let $k$ of the $N$ candidates of a mixed group be correct.
Let $\bar{\mathbf{s}}$ and $\Sigma$ be the mean and the covariance matrix of $\mathbf{s}_i$ over the correct candidates, with normalization $1/k$.
Let $m = \mathbf{w}^\top \bar{\mathbf{s}} > 0$.
Then $\eta > 0$ and
\begin{equation}
\label{eq:eta_weights}
\eta^2 = \frac{(N - k)\, m^2}{N\, \mathbf{w}^\top \Sigma\, \mathbf{w} + (N - k)\, m^2}.
\end{equation}
\end{corollary}
\begin{proof}
The rewards of the correct candidates have mean $m$ and variance $v = \mathbf{w}^\top \Sigma\, \mathbf{w}$, and every other reward is zero.
The group mean is $km/N$, so the variance of $R$ in the group is $\tfrac{k}{N}\, v + \tfrac{k}{N}\big(1 - \tfrac{k}{N}\big) m^2$.
The class means are $m$ and $0$, so the between-class variance is $\tfrac{k}{N}\big(1 - \tfrac{k}{N}\big) m^2$.
By \cref{prop:agree}, $\eta^2$ is the ratio of the second variance to the first.
The covariance of $R$ with the indicator $g$ is $\tfrac{k}{N}\big(1 - \tfrac{k}{N}\big) m$, which is positive.
\end{proof}

Equation \eqref{eq:eta_weights} yields three implications. Firstly, multiplying all weights by a positive constant does not alter the equation, indicating that only the ratios of the weights are significant. Secondly, at $\Sigma = 0$, the equation results in $\eta = 1$, corresponding to the scenario described in \cref{cor:idle}. Thirdly, for a correct candidate, $S_F$ is constrained within the interval $[\tau_F, 1]$, implying that the standard deviation of $S_F$ among correct candidates is at most $(1 - \tau_F)/2 = 0.005$. To terms of this order, the fidelity weight $w_F$ influences $m$ but not $\mathbf{w}^\top \Sigma\, \mathbf{w}$. Consequently, an increase in $w_F$ shifts $\eta$ closer to 1, aligning the rubric with a binary reward system. The cost scores differentiate correct candidates solely through $\mathbf{w}^\top \Sigma\, \mathbf{w}$.

\subsection*{Proofs}
\phantomsection\label{sec:proofs}
We formalized the statements in Lean 4 utilizing the Mathlib library, with the Lean kernel verifying each proof. The file includes \cref{lem:gate}, \cref{prop:agree} with its variance share, \cref{cor:idle}, and \cref{prop:select}. Regarding \cref{lem:silent}, it addresses a finite form that maintains the weights of the pass patterns of $N$ independent candidates and the Jensen step for a finite set of prompts. The sign of $\eta$, the identity $v = \mathbf{w}^\top \Sigma\, \mathbf{w}$. Each theorem relies solely on the three standard axioms of Lean and Mathlib. The sources employ Lean 4.36.0 and Mathlib 4.36.0 (\texttt{lean/proofs} in the code archive; Supplementary Note 8). Additionally, a test file verifies each statement on random groups to machine precision (\texttt{tests/test\_statements.py}).

\subsection*{Reward variants}
\label{sec:variants}
Each variant evaluates a candidate using the same five scores relative to the target of its specific prompt, differing in their combination methods:

\begin{itemize}
\item \textbf{Rubric}: As defined by Eq.~\eqref{eq:rubric_reward}, this is the RubriQ reward.
\item \textbf{Binary}: Represented by $g(c)$, this score is determined solely by the verifier's output: 1 for a correct circuit and 0 otherwise, serving as the control without a cost term.
\item \textbf{Scalar $T$-count}: Calculated as $g(c)\,S_T(c)$, this is a gated cost score, analogous to the sparse $1/T$ reward from the previous model evaluation, and serves as the single-cost baseline.
\item \textbf{Fidelity, no gate}: Denoted as $S_F(c)$, this score is applied to every parsed circuit without gating.
\item \textbf{Rubric without gate}: Expressed as $\sum_d w_d S_d$, this score applies to every parsed circuit.
\item \textbf{Rubric, cost scores gated}: Formulated as $w_F S_F(c) + g(c)\,[w_C S_C + w_T S_T + w_H S_H + w_E S_E]$, where fidelity is assessed for every parsed circuit, and cost scores are considered above the gate. This variant is distinct from RubriQ's reward, introduced to isolate the impact of grading fidelity below the gate. \hyperref[sec:r4]{Results} refer to this as the dense variant.
\end{itemize}

Each candidate's scores, gate result, and $T$-count are logged, enabling the offline recomputation of any variant's reward from any run.

\subsection*{Training}
\label{sec:training_loop}
In each optimization step, 8 prompts are sampled, resulting in $N = 8$ candidates per prompt, totaling 64 candidates, all processed on a single NVIDIA A100 graphics processing unit (GPU). The GRPO implementation of \texttt{trl}~0.24 is employed, utilizing the plain GRPO loss without a KL term and incorporating group-standard-deviation scaling of the advantage. A single policy update per generation ensures the precise execution of update Eq.~\eqref{eq:grpo_gradient}. The learning rate is set at $5 \times 10^{-6}$, employing cosine annealing and a warm-up ratio of 0.1, with the AdamW optimizer~\cite{loshchilov2019decoupled}, bf16 precision, and gradient checkpointing. The planned duration of the run is 1,000 steps, encompassing 64,000 candidates, with checkpoints occurring every 50 steps.

The results are reported at their most recent evaluated checkpoint. The initial model evaluation varied in four parameters. It employed an OpenQASM 2.0 prompt, evaluated 32 candidates per step, allowed up to 2,048 tokens, utilized three seeds, and incorporated the absolute $S_T$.

\subsection*{Evaluation}
\label{sec:setup}
\textbf{Training-time metrics.} For every step we log four quantities.
These are the parse rate, the pass rate (fraction of candidates with $F_{\text{avg}} \ge \tau_F$), the fraction of silent groups, and the mean $T(c)/T_0$ over admissible candidates.
A silent group is a group whose eight rewards agree to within $10^{-9}$.

\textbf{Held-out Evaluation.} For each held-out target, we sample $k = 8$ candidates using the training decoding settings. We report the parse rate, the pass rate at $\tau_F = 0.99$, the number of solved targets, and the mean $T(c)/T_0$ over admissible candidates. A target is considered solved if at least one candidate is admissible. Each evaluation consists of 64 candidates.

\textbf{Compression.} For a tool or policy output $c$ of target with reference count $T_0$, the compression is $C = T_0 / T(c)$, defined when $c$ is admissible.

\subsection*{Baselines}
\label{sec:baselines}
Each compiler runs on its own, on the expanded reference circuit of each target.
The settings are as follows.
\begin{itemize}
\item Qiskit~\cite{javadi2024quantum}: \texttt{transpile} at optimization level 3 with a fixed seed, basis $\{h, s, s^\dagger, t, t^\dagger, x, y, z, cx, cz, swap, r_z\}$.
\item TKET~\cite{sivarajah2020tketar}: \texttt{FullPeepholeOptimise}, rebased to the same basis, with implicit wire swaps made explicit.
\item PyZX~\cite{kissinger2019pyzx}: ZX-calculus \texttt{full\_reduce} with circuit extraction and basic optimization.
\item PyZX \texttt{phase\_block\_optimize} with and without pre-optimization, the latter being TODD~\cite{heyfron2018efficient}, and \texttt{full\_optimize}. These take Clifford+$T$ input, so they apply to the 4 targets without arbitrary-angle rotations.
\item BQSKit~\cite{niu2024effectiveqr}: \texttt{compile} at optimization level 2 with \texttt{max\_synthesis\_size} 3 and synthesis tolerance $10^{-8}$.
\end{itemize}
Each output is evaluated against the target unitary, with the output permutation applied when recorded by the tool, and $T$-counted using the rubric's estimator. The base model, utilizing identical prompts and lacking adapters, serves as the memorization baseline. Supplementary Note~10 provides a zero-shot reference for five additional open models employing the same prompts, parser, and verifier. On the target set, their pass rates range from 0.00 to 0.15, yielding 0 to 7 correct samples out of 64 on the held-out targets. We did not execute AlphaTensor-Quantum~\cite{ruiz2024alphatensorq}, as it optimizes a given circuit rather than constructing one from a specification, necessitating a distinct comparison protocol.

\subsection*{Experiment}
\label{sec:selection_methods}
The selection experiment implements the update of Eq.~\eqref{eq:grpo_gradient} on a policy restricted to selecting among verified circuits. Consequently, the reward represents the sole variable that varies between runs.

\textbf{Pools.} Each target is associated with a pool of candidate circuits. The correct candidates are derived from published constructions and compilers, with each candidate undergoing verification specific to its family. Circuits that do not pass the verification process are excluded, and duplicate circuits are counted only once. This process resulted in the exclusion of the Qiskit level-3 output for 10 targets and the T-par file for another 10; no search for qubit relabeling was conducted. An incorrect candidate is defined as a copy of a correct circuit with one gate either deleted or relocated to different qubits. Such candidates are retained only if they fail the verification process. Each pool contains nine incorrect candidates for every correct one, resulting in a uniform policy pass rate of 0.10. The benchmark comprises 84 targets and 9,850 candidates across five families.
\begin{itemize}
\item \textbf{Clifford} (20 targets). This task involves uniformly random Clifford operators applied to 5, 10, 20, and 50 qubits, with five instances for each qubit count. The correct candidates are derived from four synthesis methods in Qiskit: Aaronson--Gottesman, greedy, layered, and layered for linear connectivity. The verifier performs a comparison using stabilizer tableaux~\cite{aaronson2004improved}.

\item \textbf{Entanglement} (18 targets). As described in ref.~\cite{furrutter2024quantum}, the entanglement-generation task involves a target defined by a Schmidt rank vector (SRV) on 3 to 8 qubits, utilizing H and CX gates. For each qubit width, the pool contains 400 random circuits with 2 to $4n$ gates. The targets consist of three SRVs per width, which are achieved by 1--25\% of these circuits, resulting in an average uniform pass rate of 0.034. The verifier extracts the SRV from the stabilizer tableau, where qubit $i$ has a Schmidt rank of $2^{r_i - 1}$, with $r_i$ representing the rank over the two-element field GF(2) of its X and Z columns in the generator matrix.

\item \textbf{Unitary} (24 targets). This task pertains to unitary compilation as outlined in ref.~\cite{furrutter2024quantum}. It includes 15 targets on 3 qubits derived from random circuits comprising 6 to 12 gates from the set $\{$H, CX, Z, X, CCX, SWAP$\}$. Additionally, 9 targets on 4 to 6 qubits are generated from random circuits with $6n$ gates using $\{$H, S, T, CX, $R_z\}$. The correct candidates include the reference circuit, Qiskit at levels 1 and 3, and two PyZX pipelines. The verifier ensures precision with $F_{\mathrm{avg}} \ge 0.99$.

\item \textbf{$T$-count suite} (17 targets). This suite comprises arithmetic circuits on 5 to 19 qubits from the standard suite~\cite{amy2014polynomial,nam2018automated}. It includes four Toffoli and four Barenco Toffoli circuits, Mod~$5_4$, VBE-Adder$_3$, RC-Adder$_6$, CSLA-MUX$_3$, Mod-Red$_{21}$, Mod-Mult$_{55}$, and the GF($2^m$) multipliers for $m = 4, 5, 6$. The correct candidates are the input circuit, the published outputs of Nam et al.~\cite{nam2018automated} and of T-par~\cite{amy2014polynomial}, and two PyZX pipelines~\cite{kissinger2020reducing}. The verifier requires a state fidelity of at least 0.99 on each of three random product input states~\cite{burgholzer2021random}.

\item \textbf{Adders} (5 targets). The $n$-bit ripple-carry adder is implemented for $n = 4, 8, 16, 32, 64$ using $3n$ qubits, employing X, CX, and CCX gates. The constructions by ~\cite{vedral1996quantum} and ~\cite{cuccaro2004new} are identified as correct candidates. The verification process involves executing the circuit as reversible logic on 1,024 random basis inputs, assessing the sum, the carry, and the clean helper qubits. The time complexity is linear with respect to the gate count, similar to kickmix circuits~\cite{khattar2026leveraging}. The absence of failures in 1,024 inputs constrains the failure probability on a random input to below 0.3\% with 95\% confidence.
\end{itemize}

\textbf{Scoring.}The evaluator of the model evaluations assigns each candidate five scores, with reward variants being those that exceed a certain threshold. The score of the verifier, denoted as $S_F$, represents fidelity, which is the proportion of tableau rows, SRV entries, or basis inputs that are in agreement. A candidate successfully passes the gate if $S_F \ge 0.99$ for fidelity, and $S_F = 1$ in other cases. For a target where $T_0 = 0$, the score $S_T$ assumes an absolute form, equating to 1 when $T(c) = 0$.

\textbf{Policy and update.} The policy of a target is a softmax over the candidates of its pool, with logits $\theta$ that start at zero.
Policies of different targets share no parameters.
One step samples 8 targets with replacement and $N = 8$ candidates for each.
It applies $\theta \leftarrow \theta + \lambda\,\tfrac{1}{N}\sum_i \hat{A}_i \nabla_\theta \log \pi_\theta(c_i)$ with the advantages of Eq.~\eqref{eq:grpo_advantage} and $\varepsilon = 10^{-4}$.
The step size is $\lambda = 0.1$, a run has 6,000 steps, and each reward has 20 seeds per family.

\textbf{Metrics.} Every 100 steps we compute three quantities exactly from the policy of each target.
The pass rate is the probability of a correct candidate.
The gap closed is $(\bar{r} - \bar{r}_0)/(r_{\max} - \bar{r}_0)$.
Here $\bar{r}$ is the mean reward $R$ of correct candidates under the policy, and $\bar{r}_0$ is its value under the uniform policy.
$r_{\max}$ is the highest reward $R$ in the pool, where $R$ is the rubric of Eq.~\eqref{eq:rubric_reward} for every run.
A cost ratio is the mean $T$-count, depth or two-qubit gate count of correct candidates under the policy.
We divide it by the lowest value among the correct candidates of the pool.
We average the pass rate and the gap closed over targets by the arithmetic mean and the cost ratios by the geometric mean.

\textbf{Step size.} We repeated the experiment with five seeds at $\lambda = 0.5$ for 1,200 steps and at $\lambda = 0.02$ for 30,000 steps.
In the four families of the main result, the rubric closed 0.48--0.92 of the gap at $\lambda = 0.5$ and 0.99--1.00 at $\lambda = 0.02$.
The binary reward closed at most 0.04 and 0.28.
On the entanglement targets the rubric closed 0.09 and 0.15.
The gap closed thus rises as the step size falls (Supplementary Note~7).

\paragraph{Readout-error bound}
\phantomsection\label{sec:readout}
We bound the share of $D_{\mathrm{KL}}$ that readout error can cause, under two assumptions (we refer \hyperref[sec:r6]{Results}).
First, each qubit $i$ is read out independently, with flip probabilities $r_{01}^{(i)} = P(1 \mid 0)$ and $r_{10}^{(i)} = P(0 \mid 1)$.
This is the tensor-product model of readout mitigation~\cite{bravyi2021mitigating}.
Second, the QFT test circuit acts on a computational basis state, so its noiseless output $u$ is uniform on $\{0,1\}^n$.
Let $p$ be the distribution of the $n$ bits just before readout, after all gate noise.
Let $M$ be the readout channel, so that the measured distribution is $q = Mp$.

\begin{proposition}\label[proposition]{prop:readout}
Let $\Delta_i = \lvert r_{01}^{(i)} - r_{10}^{(i)} \rvert$, and measure $D_{\mathrm{KL}}$ in nats.
Under the two assumptions,
\begin{equation}
\label{eq:readout_bound}
D_{\mathrm{KL}}(q \,\|\, u) \le D_{\mathrm{KL}}(p \,\|\, u) + \sum_{i=1}^{n} \ln(1 + \Delta_i).
\end{equation}
For perfect gates, $p = u$, the bound tightens to
\begin{equation*}
D_{\mathrm{KL}}(q \,\|\, u) \le \sum_{i=1}^{n} \ln(1 + \Delta_i^2) \le n \max_i \Delta_i^2 .
\end{equation*}
\end{proposition}
\begin{proof}
We split the divergence:
\begin{equation*}
\begin{aligned}
D_{\mathrm{KL}}(Mp \,\|\, u) ={}& D_{\mathrm{KL}}(Mp \,\|\, Mu) \\
& + \mathbb{E}_{Mp}\big[\ln\big((Mu)(x)/u(x)\big)\big].
\end{aligned}
\end{equation*}
The data-processing inequality bounds the first term by $D_{\mathrm{KL}}(p \,\|\, u)$~\cite{cover2006elements}.
Readout turns a uniform bit $i$ into a bit with $P(1) = \tfrac{1}{2} + \delta_i$, where $\delta_i = (r_{01}^{(i)} - r_{10}^{(i)})/2$.
Thus $(Mu)(x)/u(x) \le \prod_i (1 + 2\lvert\delta_i\rvert) = \prod_i (1 + \Delta_i)$ for every bit string $x$, which bounds the second term.
For $p = u$, Jensen's inequality gives
\begin{equation*}
\begin{aligned}
D_{\mathrm{KL}}(Mu \,\|\, u) &\le \ln \mathbb{E}_{Mu}\big[(Mu)(x)/u(x)\big] \\
&= \sum_i \ln(1 + 4\delta_i^2),
\end{aligned}
\end{equation*}
and $4\delta_i^2 = \Delta_i^2$.
\end{proof}

A symmetric readout error, represented as $\Delta_i = 0$, does not increase $D_{\mathrm{KL}}$ beyond its value prior to readout. The bound increases linearly with $\Delta_i$ when gate noise is present and quadratically when it is absent. In the case of ideal gates, the readout error alone reaches the threshold $D_{\mathrm{KL}} = 0.05$ only if $\max_i \Delta_i \ge 0.11$. When $D_{\mathrm{KL}}$ is expressed in bits, both sides of the bound are divided by $\ln 2$, leading to the condition $\max_i \Delta_i \ge 0.09$. This proposition is not applicable if the test circuit did not operate on a computational basis state.

\subsection*{Statistics}
\label{sec:stats}
The second model evaluation has two seeds per variant and the first model evaluation three; we report every seed and give no confidence interval for $n = 2$.
Training-time curves are averages over 10 consecutive steps (640 candidates).
Held-out evaluations have $n = 64$ candidates per adapter.
All data splits use seed 42.
The mixed-group rows of \cref{tab:mechanism} come from the per-candidate reward logs of the two rubric runs.
A resumed run repeats some steps, and we count each step once, from the later segment.
We compute the advantages by Eq.~\eqref{eq:grpo_advantage} with $\varepsilon = 10^{-4}$, the value of the training code.
The cost-diversity rows count, over all eight runs, the groups with two or more correct candidates whose rubric rewards differ by more than $10^{-9}$.
The selection experiment has 20 seeds per reward and family.
We report the mean over seeds and the 95\% interval from Student's $t$ distribution with 19 degrees of freedom.

\section*{Data and code availability}
The code and the data are in one archive on Zenodo at \url{https://doi.org/10.5281/zenodo.24516153}.
The archive holds the rubric evaluator, the GRPO training pipeline, the selection experiment with its verifiers, the Lean proofs and the tests.
It also holds the hardware submission scripts and the baseline comparison script.
The target set is defined in \hyperref[sec:datasets]{Methods}, with the unitary-level hold-out (seed 42, 8 of 42 targets).
The prompt dataset, that split and the baseline outputs are in the archive.
The 1,500-circuit collection of the public version is not in the archive.
Source data for Fig.~2 and Tables~\ref{tab:mechanism} and~\ref{tab:main}, per run and seed, accompany the paper as comma-separated values (CSV) files (\texttt{results/rerun\_2026-10/source\_data} in the archive).
The reward logs and the scripts behind the mixed-group and cost-diversity rows of Table~\ref{tab:mechanism} are in \texttt{results/rerun\_2026-10/e2}.
The pools, the curves of every seed and the final values of the selection experiment (Fig.~3) are in \texttt{results/selection\_2026-11}.
The circuits of the $T$-count suite come from the PyZX repository (\url{https://github.com/zxcalc/pyzx}, \texttt{circuits/Fast}).
An interactive demonstration runs at \url{https://huggingface.co/spaces/qcsyn/rubriq-demo}.
Reproduction instructions are in Supplementary Note~6.

\section*{Acknowledgements}
This research used resources of the National Energy Research Scientific Computing Center,
 a DOE Office of Science User Facility supported by the Office of Science of the U.S. Department of Energy under Contract No. DE-AC02-05CH11231 using NERSC award NERSC DDR-ERCAP0035700.

\section*{Author contributions}
Z.G. and Z.P. designed the study. Z.G. built RubriQ and the benchmark, ran the experiments, analysed the data and wrote the manuscript. Both authors revised the manuscript and approved the final version.

\section*{Supplementary information}
Supplementary Notes 1--10, Supplementary Figs.~1--3 and Supplementary Tables~1--9 are in the ancillary file \texttt{supplementary\_information.pdf}.

\section*{Competing interests}
The authors declare no competing interests.

\bibliography{references_npj}

\end{document}